\documentclass[reprint, amsmath, amssymb, aps, twocolumn]{revtex4}

\usepackage{graphicx}
\usepackage{dcolumn}
\usepackage[arrow, matrix,curve]{xy}
\usepackage{hyperref}

\newcommand{\AHI} {\A_{\mathrm{cosm}}}
\newcommand{\GEL} {\mathcal{G}}

\newcommand{\ovl}[1] {\overline{#1}}

\newcommand{\cp} {\circ}
\newcommand{\SU} {\mathrm{SU}(2)}

\newcommand{\act} {\theta}
\newcommand{\sact} {\Theta}

\newcommand{\CAP} {C_{\mathrm{AP}}(\RR)}
\newcommand{\AP} {\mathrm{AP}}

\newcommand{\A} {\mathcal{A}}

\newcommand{\X} {\ovl{X}}
\newcommand{\x} {\ovl{x}}
\newcommand{\ux} {\underline{x}}

\newcommand{\RR} {\mathbb{R}}
\newcommand{\RB} {\mathbb{R}_{\mathrm{Bohr}}}

\newcommand{\qcR} {\underline{\RR}}
\newcommand{\qR} {\ovl{\RR}}

\newcommand{\Transl} {\Sigma}
\newcommand{\Transw} {\boldsymbol{\Sigma}'}
\newcommand{\Transwq} {\boldsymbol{\Sigma}}

\newcommand{\BP} {\boldsymbol{+}}
\newcommand{\BM} {\boldsymbol{-}}

\newcommand{\Hils} {\mathcal{H}_{\mathrm{kin}}}
\newcommand{\NB} {0_{\mathrm{Bohr}}}
\newcommand{\muB} {\mu_{\mathrm{Bohr}}}

\newcommand{\muR} {\mu_{\RR}}
\newcommand{\muRB} {\mu_{\mathrm{B}}}

\newcommand{\cC} {\mathfrak{C}}
\newcommand{\Spec} {\mathrm{Spec}}

\newcommand{\dd} {\mathrm{d}}
\newcommand{\ee} {\mathrm{e}}
\newcommand{\I} {\mathrm{i}}

\newcommand{\Borel} {\mathfrak{B}}

\newcommand{\itspace} {\vspace{0pt}}

\begin{document}

\title{Uniqueness of Measures in Loop Quantum Cosmology}

\author{Maximilian Hanusch}
 \email{maximilian.hanusch@gmx.de}
 
\affiliation{Mathematics Department, University of Paderborn}
\altaffiliation[Now at ]{Physics Department, Florida Atlantic University.}

\date{March 2, 2015}

\begin{abstract}
In \cite{UniquOfKinLQC} residual diffeomorphisms have been used to single out the standard representation of the reduced holonomy-flux algebra in homogeneous loop quantum cosmology (LQC). We show that, in the homogeneous isotropic case, unitarity of the translations w.r.t.\ the extended $\RR$-action (exponentiated reduced fluxes in the standard approach) singles out the Bohr measure on both the standard quantum configuration space $\RB$ as well as on the Fleischhack one ($\RR \sqcup \RB$). Thus, in both situations, the same condition singles out the standard kinematical Hilbert space of LQC. 
\end{abstract}

\maketitle

\section{Introduction}
\label{sec:intro}
One of the most challenging problems of modern physics is the unification of quantum mechanics and general relativity to a superordinated physical theory. Such a unified description is expected to play a role whenever massive objects are concentrated on small scales. This is the case, e.g., for Big Bang scenarios or Black Holes, so that highly symmetric configurations (as in cosmology) provide an excellent context to test any such unification.\\
\indent
 Now, instead of constructing a theory of quantum gravity from the ground, one can also quantize gravity directly. One of the most promising of these approaches is loop quantum gravity \cite{BackLA, Thiemann} (LQG), a quantization based on a Hamiltonian reformulation of general relativity. Following the lines of Ashtekar, solutions of Einstein's equations are represented by smooth connections and Dreibeins on a suitable $\SU$-principal bundle and its 3-dimensional base manifold. Then, the classical configuration space of $\SU$-connections (identified with parallel transport along embedded analytic curves) is replaced by generalized parallel transports, forming a Gelfand compactification of the classical space. Here, the main reason for switching from classical to generalized parallel transports is that on the quantum space a gauge invariant measure exists, the Ashtekar-Lewandowski one. \cite{ProjTechAL} Such gauge invariant measures are essential for the definition of a physically meaningful kinematical Hilbert spaces on which the dynamics of the quantized theory can be defined.\\
\indent 
Homogeneous isotropic loop quantum cosmology (LQC) \cite{MathStrucLQG}, a symmetry reduced version of LQG, designed to describe early universe near the Big Bang, is based on the compactification of the symmetry reduced space of homogeneous isotropic connections $\AHI\cong \RR$  on $\RR^3\times \SU$. Here, originally \cite{MathStrucLQG} only parallel transports of homogeneous isotropic connections along linear curves have been quantized, leading to the Bohr compactification (compact abelian group) $\RB$ of $\RR$. The corresponding kinematical Hilbert space $\Hils$ was chosen to be the $L^2$-one that corresponds to its Haar measure $\muB$. This Hilbert space can be identified with the Hilbert space completion of the algebra (or its uniform completion $\CAP$) generated by the characters $\chi_\lambda\colon t\mapsto \ee^{\I \lambda t}$ w.r.t.\ the inner product \footnote{Use $\langle \chi_\lambda,\chi_\mu\rangle = \delta_{\lambda \mu}=\langle \GEL(\chi_\lambda),\GEL(\chi_\mu)\rangle_{\muB}$, and that by regularity of $\muB$, the continuous functions $C(\RB)=\GEL(\CAP)$ are dense in $\Hils$. Here, $\GEL$ denotes the Gelfand isomorphism.}
\begin{align}
\label{eq:scalprod}
\langle \psi_1,\psi_2\rangle:=\lim_n\frac{1}{2n}\int_{[-n,n]}\psi_1(t)\cdot\ovl{\psi_2}(t)\:\dd t.
\end{align}
Originally, the choice of $\muB$ was rather motivated by its naturality than by any physical requirement. In Section III., however, we will show that it is the unique normalized Radon measure, invariant under the action of the exponentiated reduced fluxes (momenta) in the standard representation. This is in analogy to invariance of the Ashtekar-Lewandowski measure under the  exponentiated fluxes  in full LQG.\\
\indent
Although the dynamics of standard LQC have successfully been defined \cite{MathStrucLQG}, the connection to full LQG has remained unclear for a long time. This was basically because the Bojowald-Kastrup strategy \cite{BojoKa} for the embedding of distributional states into symmetric sectors of LQG cannot be applied in this case.  
Indeed, it was shown in \cite{Brunnhack} that $\RB$ cannot be compatibly embedded into the quantum configuration space of LQG because, in contrast to the full theory, only parallel transports along linear curves have been taken into account. Such an embedding, however, is crucial for the strategy from \cite{BojoKa}, so that, to fix this problem, in \cite{CSL} parallel transport of homogeneous isotropic connections along all embedded analytic curves have been quantized. This gives rise to the Fleischhack configuration space $\RR \sqcup \RB$ (whose compact topology will be specified below) to which the Bojowald-Kastrup embedding approach can be applied once a reasonable measure has been fixed.\\
\indent
In the following, this fixing will be done by an invariance condition which singles out the Haar measure on $\RB$ for both the standard quantum configuration space as well as for the  Fleischhack one. More precisely, we will show that invariance under the canonical extensions of the $\RR$-action 
\begin{align*}
	\theta\colon \RR \times \AHI\rightarrow \AHI, \quad (t,x)\mapsto t+x
\end{align*}
to $\RR$-actions $\Transl$ and $\Transwq$ on $\RB$ and $\RR\sqcup\RB$, respectively, singles out $\muB$ under all normalized Radon measures in both situations. Here, invariance of the measures is equivalent to unitarity of the respective translation operators on the corresponding $L^2$-Hilbert spaces which, in the standard case $\RB$, represent the exponentiated reduced fluxes.

\section{Configuration Spaces}
\label{sec:prel}
Given an $\SU$-principal fibre bundle $P$, and the respective $\SU$-connections $\A$, the quantum configuration space of LQG is given by the Gelfand spectrum of the $C^*$-algebra generated by matrix entries of the parallel transports \footnote{Actually, a parallel transport $h_\gamma(A)$ is an isomorphisms between the fibres over the start and the end point of the curve $\gamma$. But, choosing a fixed element in each fibre of $P$, we can identify such an isomorphism with an element of $\SU$.} $h_\gamma\colon \A\rightarrow \SU$ along all embedded analytic curves in the base of $P$. Indeed, by compactness of $\SU$, each such matrix entry is a bounded function on $\A$, so that the uniform closure of the $^*$-algebra generated by them exists and is $C^*$.\\
\indent
In homogeneous isotropic LQC, only parallel transports w.r.t.\ homogeneous isotropic connections are quantized. More precisely, $P$ is given by $\RR^3\times \SU$, and the quantum configuration space of interest is the Gelfand spectrum of the $C^*$-algebra generated by the restrictions of $h_\gamma$ to $\AHI\cong \RR$. Here, originally only linear curves have been used to quantize, yielding to $\CAP$; the $C^*$-algebra of almost periodic functions on $\RR$ with generators $\chi_\lambda\colon t\mapsto \ee^{\I \lambda t}$. The spectrum of $\CAP$ is the compact abelian group $\RB$, whose continuous group structure is determined by
\begin{align*}
\begin{split}
	(\psi_1 \BP \psi_2)(\chi_\lambda)&:=\psi_1(\chi_\lambda)\cdot \psi_2(\chi_\lambda)\\
	\NB(\chi_\lambda)&:=1\\
	\BM\hspace{1pt} \psi(\chi_\lambda)&:=\overline{\psi(\chi_\lambda)}
\end{split}
\end{align*}
for all $\lambda\in \RR$, $\psi, \psi_1,\psi_2\in \RB$; the bar denoting the complex conjugation. Moreover, the homogeneous isotropic connections (parametrized by $\RR$) are densely embedded into $\RB$ via the homomorphism 
\begin{align}
\label{eq:iota}
\iota\colon \RR\rightarrow \RB,\quad x\mapsto [f\mapsto f(x)]
\end{align}
for $f\in \CAP$. This means that $\iota$ is continuous and injective, the closure of $\iota(\RR)$ is $\RB$, and that we have
\begin{align}
\label{eq:homomorph}
	\iota(x+y)=\iota(x)\BP\iota(y)\qquad \forall\:x,y\in \RR.
\end{align}
Now, in contrast to the standard approach using only linear curves, in the Fleischhack one all embedded analytic curves are used to define the cosmological quantum configuration space. Thus, there the spectrum of the $C^*$-algebra generated by the restrictions $h_\gamma|_{\AHI\cong \RR}$ is considered, whereby now $\gamma$ runs over all embedded analytic curves in $\RR^3$. This $C^*$-algebra turns out \cite{CSL} to be the vector space direct sum $C_0(\RR)\oplus \CAP$. Its spectrum is homeomorphic \cite{CSL} to the space  $\RR\sqcup \RB$ equipped with the compact topology generated by the sets \footnote{$\GEL(f)$ denotes the Gelfand transform of $f$ w.r.t.\ $\Spec(\CAP)$.}   
\begin{align*}
  \begin{array}{rclcl}
    V & \!\!\!\sqcup\!\!\! & \emptyset 
    && \text{$V \subseteq \RR$ open} \\
     K^c & \!\!\!\sqcup\!\!\! & \RB
    && \text{$K \subseteq \RR$ compact} \\[-1.5pt]
     f^{-1}(U) & \!\!\!\sqcup\!\!\! & \mathcal{G}(f)^{-1}(U) 
    && \text{$U \subseteq \mathbb{C}$ open, $f \in \CAP$}.
  \end{array}
\end{align*}
In fact, the homeomorphism 
\begin{align*}
\xi\colon \RR\sqcup \RB =:\qcR\rightarrow \qR:=\Spec(C_0(\RR)\oplus\CAP) 
\end{align*}
is explicitly given by \cite{CSL} 
\begin{equation}
  \label{eq:Ksiii}
  \xi(\ux) := 
  \begin{cases} 
    \hspace{32.4pt}f\mapsto f(\ux) &\mbox{if } \ux\in \RR\\ 
    f_0\oplus f_\AP\mapsto \ux(f_\AP)  & \mbox{if } \ux\in \RB
  \end{cases}
\end{equation}
for $f_0\in C_0(\RR)$, $f_\AP\in \CAP$, and obviously we have 
\begin{align}
\label{eq:iotaxiR}
	\xi(x) = \iota'(x)\qquad \forall\: x\in \RR
\end{align} 
for $\iota'$ defined as $\iota$ in \eqref{eq:iota}, but now for $f\in C_0(\RR)\oplus \CAP$.\\
\indent
In contrast to the space $\RB$, no continuous group structure (no Haar measure) can exist on $\qcR$, but we have the splitting 
$\Borel(\qcR)=\Borel(\RR)\sqcup\Borel(\RB)$ for its Borel $\sigma$-algebra. \cite{InvConLQG} From this, it follows \cite{InvConLQG} that the finite Radon measures on $\qcR$ are exactly of the form
\begin{align*}
	(\muR\oplus\muRB)(A):=\muR(A\cap \RR) + \muRB(A\cap \RB)
\end{align*} 
for $A\in \Borel(\qcR)$ with finite Radon measures $\muR$ and $\muRB$ on $\RR$ and $\RB$, respectively.  Of course, choosing $\muR=0$ and $\muRB=\muB$, gives back the standard kinematical Hilbert space of LQC \footnote{Observe that, as in the standard case, the Hilbert space completion of $C_0(\RR)\oplus \CAP$ w.r.t.\ the inner product defined  as in \eqref{eq:scalprod} equals $L^2(\RR\sqcup \RB,0_\RR\oplus \muB)\cong \Hils$.}, and we now are going to provide a motivation for this choice.

\section{Uniqueness in Standard LQC}
 We start our investigations with the Haar measure on $\RB$ which, in particular, is translation invariant w.r.t.\ all the elements of $\iota(\RR)$. Then, if we define the action  
\begin{align*}
 \Transl \colon \RR\times \RB\rightarrow \RB,\quad (t,\psi)\mapsto \iota(t)\BP \psi,
\end{align*} 
 the family $\{\Transl^*_t\}_{t\in \RR}$ is a strongly continuous one-parameter group of unitary operators
 \begin{align*}
 \Transl_t \colon \Hils\rightarrow \Hils,\quad \varphi\mapsto \varphi \cp \Transl(t,\cdot)
\end{align*} 
for $\Hils=L^2(\RB,\muB)$. 
In fact, unitarity is clear, and for the continuity statement, we have to show that 
		\begin{align*}    	
    	\textstyle\lim_{t'\rightarrow t}\Transl_{t'}^*(\varphi)=\varphi\qquad\forall\: \varphi\in \Hils
    	\end{align*}
    	holds w.r.t.\ the $L^2$-norm $\|\cdot\|_2$. This, however, follows by a simple $\epsilon/3$-argument from denseness of $C(\RB)$ in $\Hils$ (regularity of $\muB$) and that for $f\in C(\RB)$ and $\epsilon >0$ fixed, we find $\delta>0$ with 
    	\begin{align}
    	\label{eq:toshow}
    	\begin{split}
    		 \|\Transl_t^*(f)-\Transl_{t'}^*(f)\|_\infty< \epsilon\qquad\forall\: t'\in B_\delta(t). 
    	\end{split}
    	\end{align}
    	For the last statement, observe that $\epsilon>0$ given, we find a neighbourhood $U\subseteq \RB$ of $\NB$ such that \footnote{See, e.g., 3.8 Satz in \cite{Elstrodt}.}
    	\begin{align*}
    	\psi\BM\psi'\in U\quad\Longrightarrow\quad |f(\psi)-f(\psi')|<\epsilon.
    	\end{align*}    	
    	Now, for $\delta>0$ suitably small, we have $\iota(B_\delta(0))\subseteq U$ by continuity of $\iota$, hence
    	\begin{align*}
    		\iota(t)\BM\iota(t')=\iota(t-t')\in U\qquad\forall\:t'\in B_\delta(t).
    	\end{align*}
    	This shows \eqref{eq:toshow}, because its left hand side is 
    	\begin{align*}
    		\textstyle\sup_{\psi\in \RB}|f(\psi \BP \iota(t))-f(\psi\BP\iota(t'))|.
    	\end{align*}
		We now claim that the Haar measure is the only normalized \footnote{By locally finiteness, each Borel measure on a compact space is finite. Thus, the assumption that $\muB$ is normalized (up to positive scalings) just excludes $\muRB=0$.} Radon measure $\muRB$ on $\RB$ for which the $\Transl_t^*$ are unitary operators on $L^2(\RB,\muRB)$. \\
\indent
		To this end, first observe that this condition is equivalent to requiring that for each compact $K\subseteq \RB$
	\begin{align}
	\label{eq:transl}
		\muRB(\iota(t)\BP K)=\muRB(K)\qquad\forall\:t\in \RR,
	\end{align}			
 	holds. In fact, by inner regularity of $\muRB$, \eqref{eq:transl} holds for compact sets iff it holds for Borel sets, so that unitarity of $\Transl_t$ is immediate from the general transformation formula. \\
\indent
 	Conversely, if unitarity holds, we have
 	\begin{align}
 	\label{eq:iotainv}
 	\begin{split}
 	\muRB(K)&=\textstyle\int_{\RB} |\chi_K|^2\:\dd\muRB
 	= \langle\chi_K,\chi_K\rangle\\
 	&=\langle\chi_{\iota(t)\BP K},\chi_{\iota(t)\BP K}\rangle
 	=\muRB(\iota(t)\BP K)
 	\end{split}
 	\end{align}
 	for all $t\in \RR$ and $\chi_K$ the characteristic function that corresponds to $K$. We will show that this implies 
 	\begin{align*}
 	\muRB(K)=\muRB(\psi\BP K)\qquad \forall\: \psi\in \RB
 	\end{align*} 
 	for all compact $K\subseteq\RB$, hence, by inner regularity, even for each Borel set. \\
\indent
 	Now, $\muRB$ is outer regular as it is inner regular and finite. Thus, for each $\epsilon>0$, we find an open neighbourhood $U$ of $K$ with $\muRB(U)\BM\muRB(K)<\epsilon$. The claim now follows easily as by continuity of $\BP$ and compactness of $K$, we find $V\subseteq \RB$ open with $\NB\in V$ and $V\BP K\subseteq U$. In fact, by denseness of $\iota(\RR)$, there is $t\in \RR$ with $\psi\BM\iota(t)\in V$, hence 
\begin{align*}
 \muRB(\psi \BP K)\BM\muRB(K)&\stackrel{\eqref{eq:iotainv}}{=}\mu\big(\psi\BM\iota(t) \BP K\big) \BM\muRB(K)\\\
 &\leq \muRB\big(U\big) \BM\muRB(K)<\epsilon.
\end{align*}
Thus, $\muRB(\psi \BP K)\leq \muRB(K)$ as $\epsilon$ was arbitrary., and since 
 	 $\psi$ was arbitrary as well, we can apply the same argument to $K':=\psi \BP K$ and $\psi':=\BM\hspace{1pt}\psi$, providing us with 
\begin{align*}
	\muRB(K)=\muRB(\psi'\BP K')\leq \muRB(K')=\muRB(\psi\BP K).
\end{align*}

\section{The Fleischhack Configuration Space}
\label{sec:prel}
We start with a short excursus into extensions of group actions, allowing us to carry over the above uniqueness result to the Fleischhack configuration space $\RR\sqcup \RB$. 
First, recall that \cite{InvConLQG}, given a left action $\act\colon G\times X\rightarrow X$ and a unital $C^*$-subalgebra $\cC$ of the bounded functions on $X$ with 
\begin{align}
\label{eq:inv}
	\theta_g^*(\cC)\subseteq \cC\qquad\forall\:g\in G,
\end{align}
the left action
\begin{align*}
	\sact\colon G\times \Spec(\cC)\rightarrow \Spec(\cC),\quad (g,\x)\mapsto [f\mapsto \x(\act_g^*f)]
\end{align*}
is the unique one, such that for each $g\in G$,
\begingroup 
\setlength{\leftmargini}{16pt}
\begin{enumerate}
\item
	$\sact_g$ is continuous, and
\item
\itspace
	$\sact_g\cp \iota = \iota \cp \act_g$\: holds.
\end{enumerate}
\endgroup
\noindent
Here, Condition 2. just means that
    \begin{center}
		\makebox[70pt]{
			\begin{xy}
				\xymatrix{
					\X\ar@{<-}[d]^-{\iota} \ar@{->}[r]^-{\Theta_g}  &  	\X\ar@{<-}[d]^-{{\iota}}   \\
					X\ar@{->}[r]^-{\theta_g} & X  
				}
			\end{xy}
		}
\end{center}
is commutative for each $g\in G$, i.e., that $\Theta$ extends $\theta$ in the canonical way. The action $\Theta$ is continuous iff $g\mapsto \theta^*_g f\in \cC$ is continuous for $f$ running over some set of generators of $\cC$. \footnote{Continuity of $\Theta$ will not be relevant for our further considerations. Anyhow, since the extension we are going to investigate in the following will have this property, we decided to remark on this fact.}\\
\indent
Now, let $G,X=\RR$, $\cC=\CAP$ and
\begin{align*}
 \theta\colon \RR \times \RR \rightarrow \RR,\quad (t,x)\mapsto t+x
\end{align*}
the additive action. Then, 
\begin{align*} 
 \theta_t^*\chi_\lambda=\ee^{\I \lambda t}\cdot\chi_{\lambda}\qquad \forall\: t\in \RR,\quad\forall \:\lambda\in \RR, 
\end{align*} 
 hence $\theta_t^*(\CAP)\subseteq \CAP$ as $\theta_t^*$ is an isometry. Thus, $\theta$ extends uniquely to an action $\Theta\colon \RR\times \RB\rightarrow \RB$, necessarily equal to $\Transl$, because
\begingroup 
\setlength{\leftmargini}{16pt}
\begin{enumerate}
\item
	$\Transl_t(\psi)=\iota(t)\BP \psi$ is continuous as $\BP$ and $\iota$ are,
\item
\itspace
	$(\Transl_t\cp \iota)(x)=\iota(t)\BP\iota(x)\stackrel{\eqref{eq:homomorph}}{=}\iota(t+x)=\iota\cp \theta_t(x)$ holds.
\end{enumerate}
\endgroup
\noindent
In particular, $\Transl$ is continuous, as continuity of $t\mapsto \theta_t^*\chi_\lambda$ is clear from 
\begin{align*}
	\|\hspace{1pt}\theta_t^*\chi_\lambda-\theta_{t'}^*\chi_\lambda\|_\infty =\big|\ee^{\I \lambda t}-\ee^{\I \lambda t'}\big|.
\end{align*}
Now, to provide an analogous $\RR$ action on $\qR$, we first define ($\iota\colon \RR\rightarrow \RB$ still given by \eqref{eq:iota})
\begin{align*}
  \Transwq(t,\ux) := 
  \begin{cases} 
    \hspace{11.8pt}t+\ux &\mbox{if } \ux\in \RR\\ 
    \iota(t) \BP \ux & \mbox{if } \ux\in \RB,
  \end{cases}
\end{align*}
and carry it over to $\qR$ via $\xi$, i.e., we define
\begin{align*}
\Transw(t,\x):=(\xi\cp \Transwq)(t,\xi^{-1}(\x)). 	
\end{align*}

We now show that $\Transw$ equals the unique extension $\Theta$ of $\theta$ to $\qR$. This can be proven \cite{MAX} by means of nets, but it is much easier to use $\theta_t^*$-invariance of $C_0(\RR)$ and $\CAP$. \footnote{During the preparation of this article, we were informed that Christian Fleischhack has found this simplification independently. \cite{UEBER}}
Observe that $\Theta$ indeed exists, because \eqref{eq:inv} holds for $C_0(\RR)\oplus \CAP$ as well. It is even continuous, because $t\mapsto \theta_t^* f_0$ is continuous for each $f_0\in C_0(\RR)$, which easily follows from equicontinuity of $f_0|_K$ for each compact $K\subseteq \RR$. 

Now, for $\x:=\xi(x)$ with $x\in \RR$, we have
\begin{align*}
	\Transw_t(\x)(f)&\stackrel{\phantom{\eqref{eq:iotaxiR}}}{=}\xi(\iota(t)\BP \iota(x))(f)
	\stackrel{\eqref{eq:homomorph}}{=}\xi(\iota(t+x))(f)\\[-4pt]
	&\stackrel{\phantom{\eqref{eq:iotaxiR}}}{=}f(t+x)=(\theta_t^*f)(x)=\iota'_x(\theta_t^*f)\\[-4pt]
	&\stackrel{\eqref{eq:iotaxiR}}{=}\xi(x)(\theta_t^*f)=\x(\theta_t^*f)=\Theta_t(\x)(f)
\end{align*} 
for all $f\in C_0(\RR)\oplus \CAP$, whereby the last step is just due to the definition of $\Theta$. For $\x:=\xi(\psi)$ with $\psi \in \RB$, we calculate
\begin{align*}
	\Transw_t(\x)(f_0\oplus f_\AP)&=\xi(\iota(t)\BP\psi)(f_0\oplus f_\AP)\\[-4pt]
	&=(\iota(t)\BP\psi)(f_\AP)
	\stackrel{(*)}{=}\psi(\theta_t^*f_\AP)\\[1pt]
	&=\x(\theta_t^*f_\AP)=\x(\theta_t^*(f_0\oplus f_\AP))\\
	&=\Theta_t(\x)(f_0\oplus f_\AP),
\end{align*} 
whereby $(*)$ holds because 
\begin{align*}
(\iota(t)\BP\psi)(\chi_\lambda)&=\iota(t)(\chi_\lambda)\cdot\psi(\chi_\lambda)\\
&=\psi(\chi_\lambda(t)\cdot \chi_\lambda)
=\psi(\theta_t^*\chi_\lambda)
\end{align*}
for all $\lambda\in \RR$, and since the characters $\chi_\lambda$ generate $\CAP$. In the fourth and in the fifth step, we have used $\theta_t^*$-invariance of $\CAP$ and $C_0(\RR)$, respectively.\\
\indent
Now, if $\mu$ is a normalized Radon measure on $\qR$, the same arguments as in the previous section show that unitarity of the operators 
\begin{align*}
	\Transw_t\colon L^2(\qR,\mu)\rightarrow L^2(\qR,\mu),\quad \phi\mapsto \phi\cp\Transw(t,\cdot)
\end{align*} 
is equivalent to $\Transw$-invariance of $\mu$. This, in turn, is equivalent to $\Transwq$-invariance of the push forward of $\mu$ by $\xi^{-1}$, which (as we have learned in the end of Section II) is of the form $\muR\oplus\muRB$ for some finite Radon measures $\muR$ and $\muRB$ on $\RR$ and $\RB$, respectively. If $\muR$ would not be zero, by inner regularity, we would find some compact $K\subseteq \RR$ with $\muR(K)>0$, hence $\muR(\RR)=\infty$ by translation invariance and $\sigma$-additivity. Thus, $\muRB$ is normalized  and, by the definition of $\Transwq$, invariant under the translations $\Transl_t$, hence, equal to $\muB$ by Section III.

\begin{acknowledgments}
The author thanks Jonathan Engle, Christian Fleisch-hack and Benjamin Schwarz for their helpful comments on drafts of the present article.
\end{acknowledgments}

\end{document}